\documentclass[aps,prb,twocolumn,superscriptaddress,10pt,english]{revtex4-2}
\usepackage{babel}
\usepackage{amsmath}
\usepackage{graphicx}
\usepackage{amssymb}
\usepackage{xfrac}
\usepackage{txfonts}
\usepackage{color}
\usepackage[dvipsnames]{xcolor}
\usepackage{hyperref}

\hypersetup{
    colorlinks,
    linkcolor={blue},
    citecolor={blue},
    urlcolor={blue}
}

\newcommand{\aio}{$A_2$IrO$_3$}
\newcommand{\rucl}{$\alpha$-RuCl$_3$}

\newcommand{\cac}{\chi'_{\rm ac}}
\newcommand{\TN}{T_{\rm N}}

\newcommand{\zza}{zz1}
\newcommand{\zzb}{zz2}
\newcommand{\zzthree}{3f-zz}
\newcommand{\zzsix}{6f-zz}

\newcommand{\llangle}{\langle\!\langle}
\newcommand{\rrangle}{\rangle\!\rangle}

\newcommand{\lllangle}{\langle\!\langle\!\langle}
\newcommand{\rrrangle}{\rangle\!\rangle\!\rangle}

\begin{document}

\title{Field-induced intermediate ordered phase and anisotropic interlayer interactions in \texorpdfstring{$\alpha$-RuCl$_3$}{alpha-RuCl3}}

\author{C.~Balz}
\affiliation{Neutron Scattering Division, Oak Ridge National Laboratory, Oak Ridge, TN 37831, U.S.A.}
\affiliation{ISIS Neutron and Muon Source, STFC Rutherford Appleton Laboratory, Didcot OX11 0QX, UK}

\author{L.~Janssen}
\affiliation{Institut f\"ur Theoretische Physik and W\"urzburg-Dresden Cluster of Excellence ct.qmat, Technische Universit\"at Dresden,
01062 Dresden, Germany}

\author{P.~Lampen-Kelley}
\affiliation{Department of Materials Science and Engineering, University of Tennessee, Knoxville, TN 37996, U.S.A.}
\affiliation{Materials Science and Technology Division, Oak Ridge National Laboratory, Oak Ridge, TN 37831, U.S.A.}

\author{A.~Banerjee}
\affiliation{Neutron Scattering Division, Oak Ridge National Laboratory, Oak Ridge, TN 37831, U.S.A.}
\affiliation{Department of Physics and Astronomy, Purdue University, West Lafayette IN, 47906, U.S.A.}

\author{Y.~H.~Liu}
\affiliation{Neutron Scattering Division, Oak Ridge National Laboratory, Oak Ridge, TN 37831, U.S.A.}

\author{J.-Q.~Yan}
\affiliation{Materials Science and Technology Division, Oak Ridge National Laboratory, Oak Ridge, TN 37831, U.S.A.}

\author{D.~G.~Mandrus}
\affiliation{Department of Materials Science and Engineering, University of Tennessee, Knoxville, TN 37996, U.S.A.}
\affiliation{Materials Science and Technology Division, Oak Ridge National Laboratory, Oak Ridge, TN 37831, U.S.A.}

\author{M.~Vojta}
\affiliation{Institut f\"ur Theoretische Physik and W\"urzburg-Dresden Cluster of Excellence ct.qmat, Technische Universit\"at Dresden,
01062 Dresden, Germany}

\author{S.~E.~Nagler}
\affiliation{Neutron Scattering Division, Oak Ridge National Laboratory, Oak Ridge, TN 37831, U.S.A.}

\date{\today}

\begin{abstract}
In $\alpha$-RuCl$_3$, an external magnetic field applied within the honeycomb plane can induce a transition from a magnetically ordered state to a disordered state that is potentially related to the Kitaev quantum spin liquid. In zero field, single crystals with minimal stacking faults display a low-temperature state with in-plane zigzag antiferromagnetic order and a three-layer periodicity in the direction perpendicular to the honeycomb planes. Here, we present angle-dependent magnetization, ac susceptibility, and thermal transport data that demonstrate the presence of an additional intermediate-field ordered state at fields below the transition to the disordered phase. Neutron diffraction results show that the magnetic structure in this phase is characterized by a six-layer periodicity in the direction perpendicular to the honeycomb planes. Theoretically, the intermediate ordered phase can be accounted for by including spin-anisotropic couplings between the layers in a three-dimensional spin model. Together, this demonstrates the importance of interlayer exchange interactions in $\alpha$-RuCl$_3$.
\end{abstract}

\maketitle


\section{Introduction}

Frustrated magnets with strong spin-orbit coupling have attracted great interest, largely because of the possibility that they may provide realizations of quantum spin liquids (QSLs): highly-entangled topological states of matter with fractionalized excitations and emergent gauge fields. The Kitaev model on the honeycomb lattice \cite{Kit06} is a unique and solvable example, in which spin-flip excitations fractionalize into itinerant Majorana fermions and Ising gauge-field excitations.

The search for realizations of the Kitaev model has uncovered a number of insulating honeycomb-lattice magnets, in which strong spin-orbit coupling generates $J_{\rm eff}=1/2$ local moments subject to bond-dependent Ising interactions \cite{Jac09,Cha10,Tak19,janssen2019}. These include the stoichiometric crystalline materials \aio\ ($A = \mathrm{Na}, \mathrm{Li}$) and {\rucl}; however, antiferromagnetic long-range order is realized at low temperatures in these materials. Among them, {\rucl} has attracted immense attention \cite{Plu14,Sea15,Ban16} for two reasons:
(i) Spectroscopic experiments have detected \cite{Ban16,Ban17,Ban18,Bal19} clear signatures of fractionalized excitations over a significant range of energies, which have been interpreted in terms of proximate spin-liquid behavior \cite{Goh17}.
(ii) Magnetic fields applied in the honeycomb plane suppress magnetic order, leading to a spin-liquid-like state, the precise nature of which is under debate \cite{Wol17,Sea17,Bae17,Zhe17,Lea17,Win18,Hen18,Yokoi20,Gas20,Bac20,Chern20,janssen2019}. In fact, the overall temperature-magnetic field ($T$-$B$) phase diagram of {\rucl} is currently under intense scrutiny.  In zero field, single crystals with minimal stacking faults show a transition near $\TN = 7$\,K to a low-temperature ordered phase that has a zigzag antiferromagnetic (AFM) structure in a single honeycomb plane, with a three-layer periodicity perpendicular to the planes \cite{Ban16}. Some recent experimental results show that at low temperatures there is evidence for a field-induced transition to an additional ordered state before the zigzag-ordered phase is suppressed. Preliminary evidence for this was seen in ac susceptibility measurements \cite{Ban18}, and the thermodynamic nature of the transition was confirmed by heat capacity \cite{Tan20}, magnetocaloric effect (MCE) \cite{Bal19}, and magnetic Gr\"uneisen parameter \cite{Bac20} data.

Questions also remain about the full $T$-$B$ phase diagram at higher fields. The reported quantized thermal Hall conductivity at fields above the disorder threshold \cite{Kas18,Yokoi20} suggests the presence of an additional topological phase transition at a second, higher field. This appears consistent with reported MCE and inelastic neutron scattering measurements \cite{Bal19}. However, recent Raman \cite{Sah19,Wul19}, terahertz \cite{Wang17}, and electron spin resonance spectroscopy \cite{Pon20}, as well as Gr\"uneisen parameter measurements \cite{Bac20}, do not show a clear signature of such a transition. On the theoretical front, there is still considerable discussion about the appropriate microscopic Hamiltonian describing the magnetism of {\rucl} \cite{Kim11,Cha13,Rau14,Per14,Rou15,Win16,Win17,Jan17,Lam18,Suz18,Jan20,Mak20}. Clearly, the correct Hamiltonian must account for all of the experimentally observed phases and transitions.

\begin{figure}
\includegraphics[width=\columnwidth]{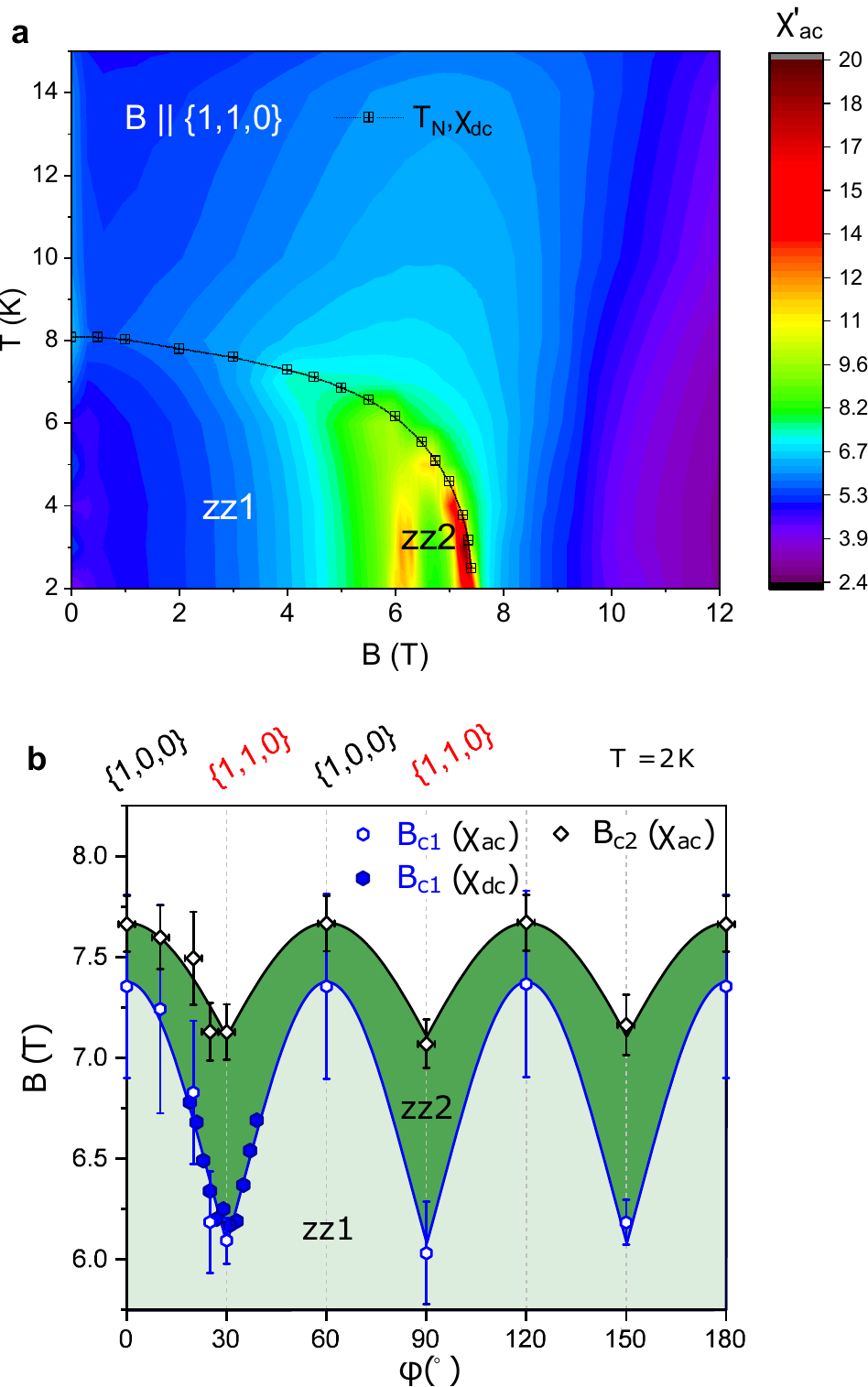}
\caption{(a) Temperature-magnetic field phase diagram of {\rucl} for in-plane fields perpendicular to Ru-Ru bonds (see Fig.~\ref{fig:honeycomb}), as constructed from ac susceptibility measurements. The N\'eel temperature as extracted from dc susceptibility measurements \cite{Ban18} is shown as black squares for comparison. The dome-shaped intermediate ordered phase {\zzb} occurs between the low-field ordered phase  {\zza} and the disordered phase at higher fields.
(b) Phase diagram of {\rucl} at $T=2$\,K as function of in-plane angle and magnetic field from ac and dc susceptibility measurements. The lines are a guide to the eye.
}
\label{fig:phase_diagrams}
\end{figure}

In this paper, multiple experimental probes are used to investigate the intermediate-field ordered phase \cite{prev_note}. Orientation-dependent magnetization and susceptibility measurements map the phase diagram as a function of magnetic field direction, strength, and temperature; see Fig.~\ref{fig:phase_diagrams}. Neutron-diffraction measurements show that the intermediate-field ordered state features a periodicity in the direction perpendicular to the honeycomb plane that is different from that of the low-field ordered state, implying that interlayer exchange interactions must be accounted for in order to understand the transition between these two states.  To that end, an effective spin Hamiltonian modeling these interactions is introduced, and is shown to describe well the field-induced transition between the two zigzag phases for appropriately chosen model parameters.

The rest of this paper is organized as follows:
In Sec.~\ref{sec:expdet}, we provide experimental details about the measurements performed. The presentation of the experimental results starts with the bulk properties in Sec.~\ref{sec:bulk}, before the neutron diffraction is addressed in Sec.~\ref{sec:neutron}. In Sec.~\ref{sec:structure_factor}, we present a modeling of the magnetic structure factor. Section~\ref{sec:model} discusses the properties of a three-dimensional (3D) spin model in an in-plane magnetic field, which is shown to reproduce the key features of the experiment.
The paper ends with a discussion in Sec.~\ref{sec:disc}.


\section{Experimental methods}
\label{sec:expdet}

\begin{figure}
\includegraphics[width=0.75\columnwidth]{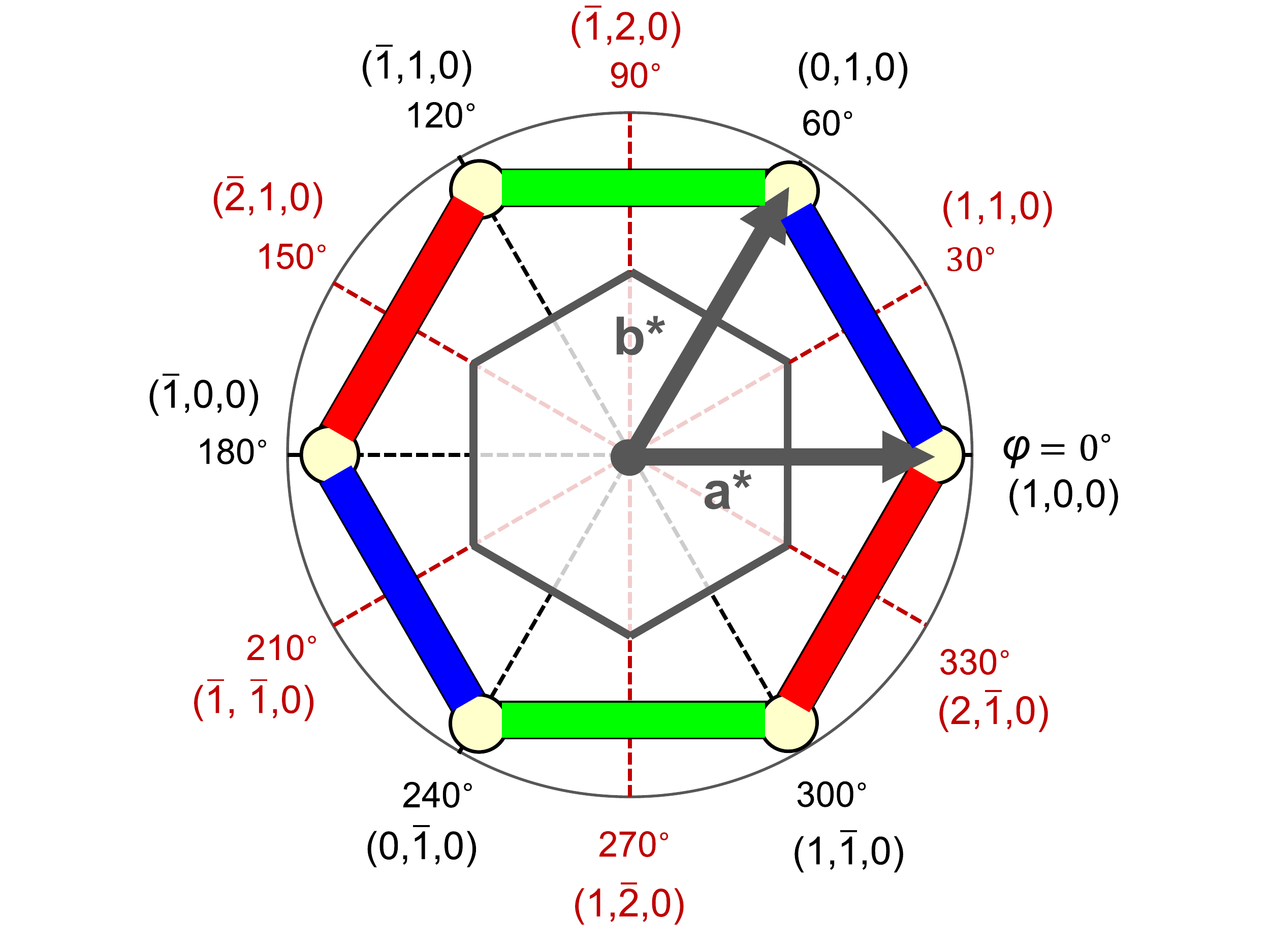}
\caption{Definition of the in-plane angle $\varphi$ within the first Brillouin zone of the two-dimensional (2D) reciprocal lattice (gray hexagon) and with respect to the real-space orientation of the Ru-Ru bonds (red/blue/green). Directions that are equivalent to (1,0,0) (black) and (1,1,0) (red) correspond to angles $\varphi \equiv 0^\circ \bmod 60^\circ$ and $\varphi \equiv 30^\circ \bmod 60^\circ$, respectively. $\mathbf{a}^*$ and $\mathbf{b}^*$ denote the reciprocal lattice vectors in the $R\bar{3}$ structure.
}
\label{fig:honeycomb}
\end{figure}

Susceptibility measurements were performed on {\rucl} single crystals prepared by a vapor transport method described elsewhere \cite{Ban17} and oriented by X-ray Laue diffraction using a conventional $R\bar{3}$ unit cell, see definition in Fig.~\ref{fig:honeycomb}. Angle-resolved dc magnetization measurements were collected using a sample rotation stage in a 7\,T SQUID magnetometer. dc magnetization, ac susceptibility, and thermal-transport measurements were performed up to 14\,T at various fixed angles in a Physical Property Measurement System (Quantum Design).

Neutron diffraction measurements were performed on the CORELLI instrument at the Spallation Neutron Source using an 8\,T vertical-field cryomagnet. CORELLI is a time-of-flight instrument with a pseudo-statistical chopper, which separates the elastic contribution \cite{Ros08}. For this experiment, a 2\,g single crystal of {\rucl} studied earlier \cite{Bal19} was mounted on an aluminum sample holder and aligned with the $(H,0,L)$ plane in the horizontal scattering plane. This way, the magnetic field of the 8\,T vertical-field cryomagnet is aligned parallel to a $\{1,1,0\}$-equivalent direction. The crystal was rotated through $360^\circ$ in steps of $4^\circ$. Large vertical detector coverage at CORELLI allows access to the full set of magnetic Bragg peaks in the honeycomb 2D Brillouin zone, which is oriented vertically in this configuration. For all data shown, a measurement at 8\,T is subtracted as a background after it was confirmed that no elastic magnetic intensity remained at this field strength. The data was reduced using Mantid \cite{Arn14}.

\begin{figure*}
\includegraphics[width=\textwidth]{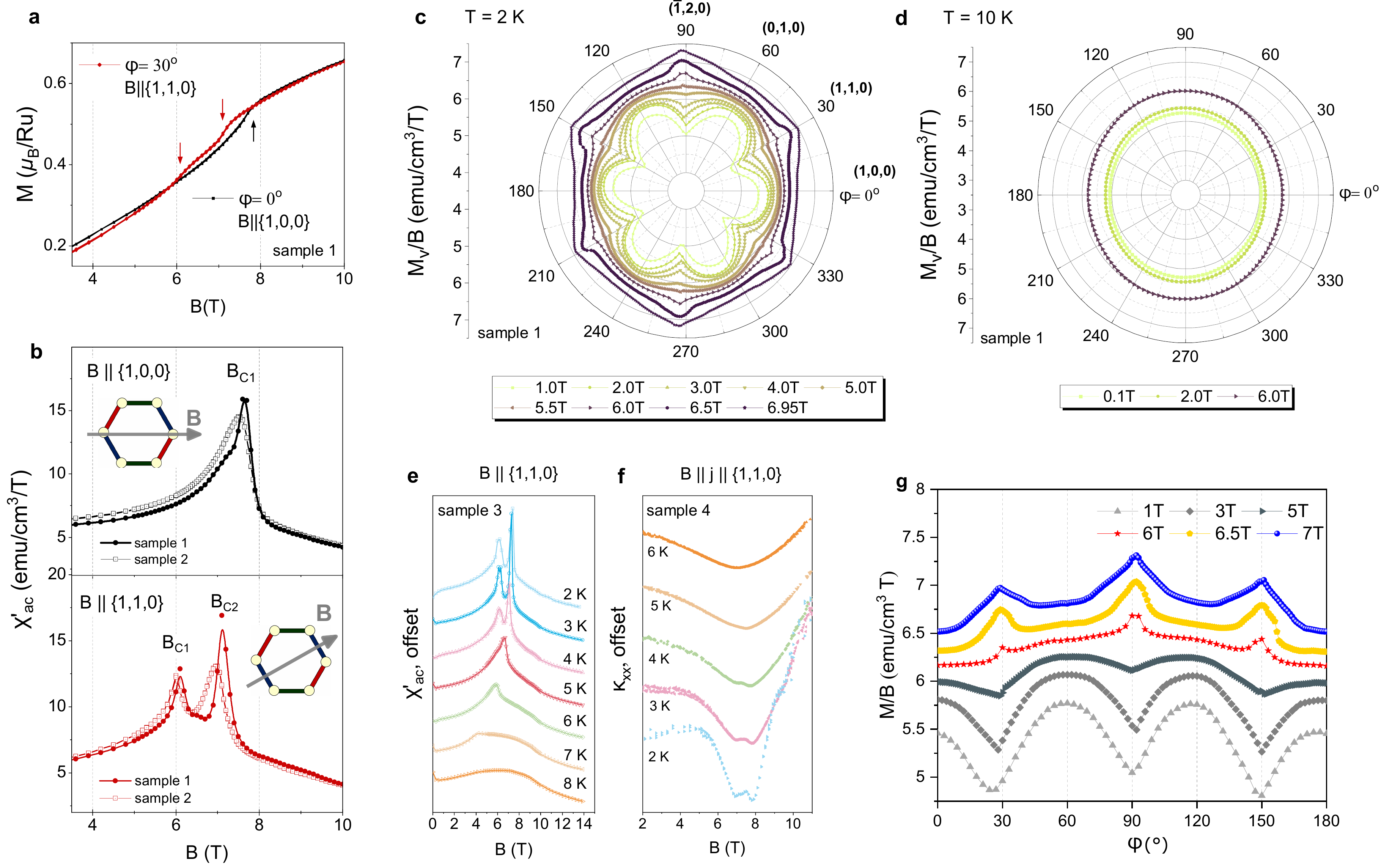}
\caption{
(a) Field-dependent magnetization for different in-plane field directions $\varphi$ at $T=2$\,K. Arrows indicate kinks in the magnetization.
(b) Real part of the ac susceptibility $\cac$ as a function of dc magnetic field for in-plane field directions $\mathbf B\parallel\{1,0,0\}$ ($\varphi \equiv 0^\circ \bmod 60^\circ$) (upper panel) and  $\mathbf B\parallel\{1,1,0\}$ ($\varphi \equiv 30^\circ \bmod 60^\circ$) (lower panel), at $T=2$\,K. The frequency of the 1\,mT ac field is 1\,kHz.
(c) Polar plot of the angular dependence of the dc magnetization ($M/B$) plotted for various field strengths at $T=2$\,K. A sixfold oscillation as a function of $\varphi$ is visible. The maxima and minima are reversed around $6$\,T.
(d) The angle dependence of the magnetization above $\TN$ at $T=10$\,K.
(e) $\cac$ at various fixed temperatures showing two anomalies as a function of $\{1,1,0\}$ magnetic field strength for $T\lesssim 4$\,K.
(f) Thermal conductivity at various fixed temperatures as a function of $\{1,1,0\}$ magnetic field strength. The curves in (e) and (f) are offset for clarity.
(g) Linear plot of the angle dependence of the magnetization ($M/B$) at various fields, for $T=2$\,K. The data is the same as that plotted in (c).  The exchange of the minima and maxima between 5 and 6\,T is clearly visible in this plot.
}
\label{fig:bulk_properties}
\end{figure*}


\section{Bulk properties}
\label{sec:bulk}

\begin{figure*}
\includegraphics[width=0.8\textwidth]{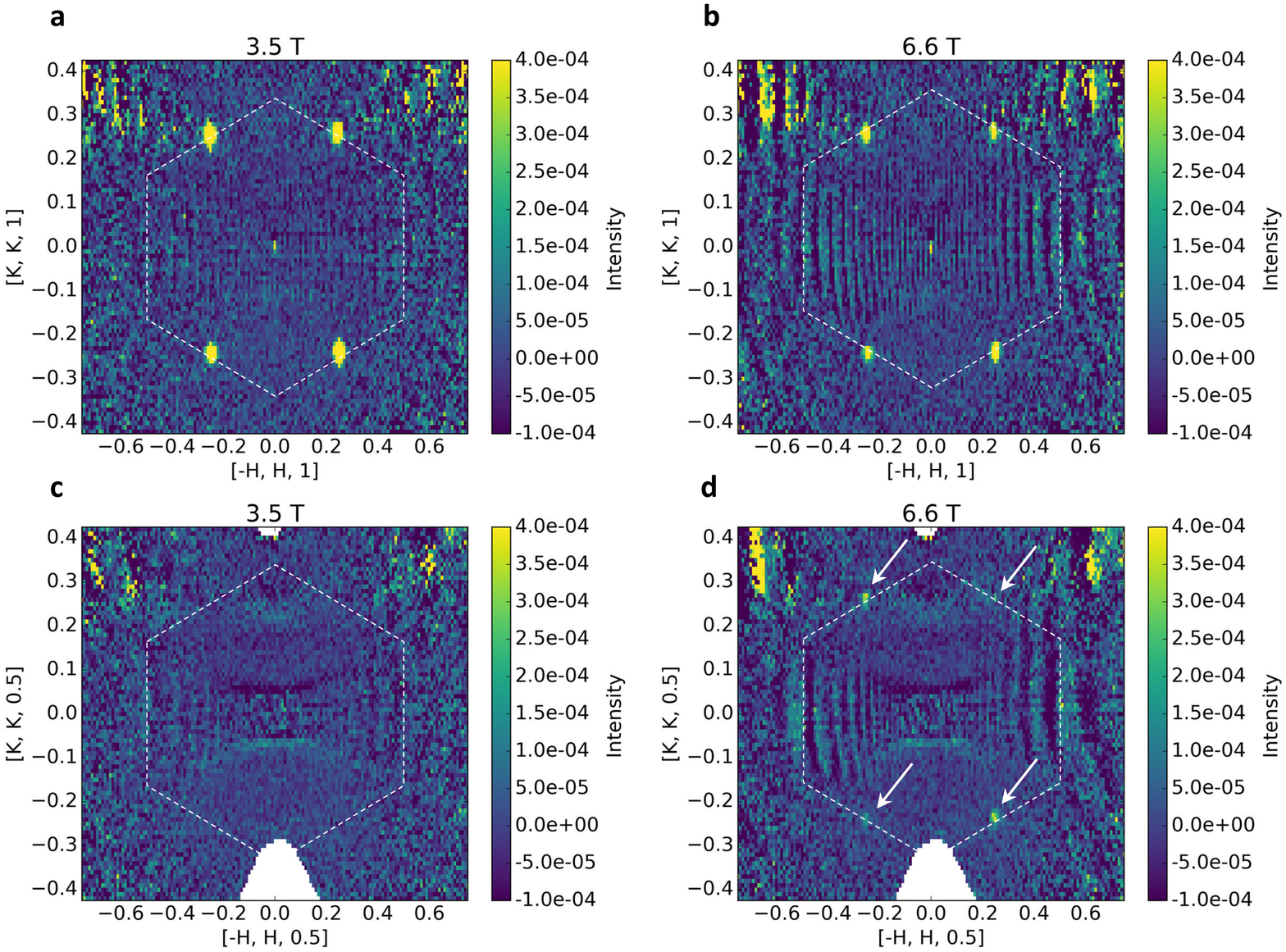}
\caption{Top row: Neutron diffraction intensities at $|L|=1$ for {(a)} $B=3.5$\,T and {(b)} $B=6.6$\,T. Magnetic Bragg peaks are visible at the $\mathbf M$ points $(1/2,0,1)$, $(-1/2,0,1)$, $(0,-1/2,1)$, and $(0,1/2,1)$. Bottom row: Intensities at $|L|=0.5$ for {(c)} $B=3.5$\,T and {(d)} $B=6.6$\,T. Additional weak magnetic Bragg peaks appear at the $\mathbf M$ points in the 6.6\,T data. The perpendicular integration range in is $\Delta L=\pm0.025$\,rlu and the data has been averaged over positive and negative $L$. The first Brillouin zone is indicated by the dashed hexagon and the arrows in {(d)} point to the weak intensity observed at at the $\mathbf M$ points. The stripy intensity appearing in the upper corners is spurious and caused by imperfect background subtraction for larger wavevectors.
}
\label{fig:diff_slices}
\end{figure*}

When a magnetic field $\mathbf B$ is applied parallel to a Ru-Ru bond [corresponding to one of the symmetry-equivalent $(1,0,0)$, $(0,1,0)$, or $(-1,1,0)$ directions, see Fig.~\ref{fig:honeycomb}], the magnetization at 2\,K shows a single kink at $\simeq 7.6$\,T in the vicinity of the well-documented field-induced suppression of the zigzag ordered phase \cite{Wol17,Sea17,Bae17,Zhe17,Lea17,Win18,Hen18}, see Fig.~\ref{fig:bulk_properties}{(a)}. Minor variation between samples is present, as shown in Fig.~\ref{fig:bulk_properties}{(b)}. Rotating the magnetic field perpendicular to a bond, i.e., along a $\{1,1,0\}$-equivalent direction, reveals a second feature near 6\,T, well below the purported transition into the field-induced disordered phase. The anisotropy of the critical fields within the honeycomb plane is clearly visible in ac susceptibility $\cac$ measurements, Fig.~\ref{fig:bulk_properties}{(b)}. Two well-separated anomalies in $\cac$ at $B_\mathrm{c1} \simeq 6$\,T and $B_\mathrm{c2} \simeq 7-7.3$\,T as a function of $\{1,1,0\}$ field strength converge and shift slightly higher to $B_\mathrm{c2} \simeq 7.6$\,T in a $\{1,0,0\}$ field. This behavior repeats every $60^{\circ}$, consistent with the symmetry of the honeycomb lattice, and has been reproduced in a number of samples.

Figs.~\ref{fig:bulk_properties}{(c)} and {(d)} show the angle dependence of the magnetization obtained via sample rotation in a field up to 7\,T at 2\,K and 10\,K, respectively. Here, $\varphi$ is the angle between the magnetic field $\mathbf B$ and the reciprocal lattice vector $\mathbf{a^*}$, cf.\ Fig.~\ref{fig:honeycomb}. At moderate fields $\gtrsim 1$\,T, angle-resolved magnetization below $\TN=7$\,K exhibits a sixfold symmetry with maxima at $\varphi \equiv 0^\circ \bmod 60^\circ$, where the field coincides with a bond-parallel $\{1,0,0\}$ direction. The amplitude of this oscillation decreases with increasing field. At elevated fields $\gtrsim 6$\,T, a distinct set of maxima appear in a narrow range of $\varphi$ around the $\{1,1,0\}$ directions $\varphi \equiv 30^\circ \bmod 60^\circ$, also clearly visible in Fig.~\ref{fig:bulk_properties}{(g)}. Above $\TN$ the oscillation in the angle dependence of the magnetization disappears (Fig.~\ref{fig:bulk_properties}{(d)}.

The double-peak behavior in $\cac (B)$ in a $\{1,1,0\}$ magnetic field emerges several Kelvin below the N\'eel transition, becoming distinct only for $T\lesssim 4$\,K, Fig.~\ref{fig:bulk_properties}{(e)}. Thermal-conductivity measurements exhibit consistent behavior, as shown in Fig.~\ref{fig:bulk_properties}{(f)}. A minimum in $\kappa_{xx} (B)$ marking the critical field for the suppression of the zigzag order in \rucl\ has been previously reported \cite{Lea17,Hen18,Yu18}. With $\mathbf B \parallel\{1,1,0\}$ this feature splits into two distinct minima below 5\,K. We note that the $\kappa_{xx} (B)$ minima near 7 and 7.8\,T are larger than the analogous critical fields in susceptibility data; the detailed field-dependence of magnetic contributions to phonon scattering and $\kappa_{xx}$ enhancement across the two transitions are not well understood and likely play a role in the discrepancy.

\begin{figure*}
\includegraphics[width=0.8\textwidth]{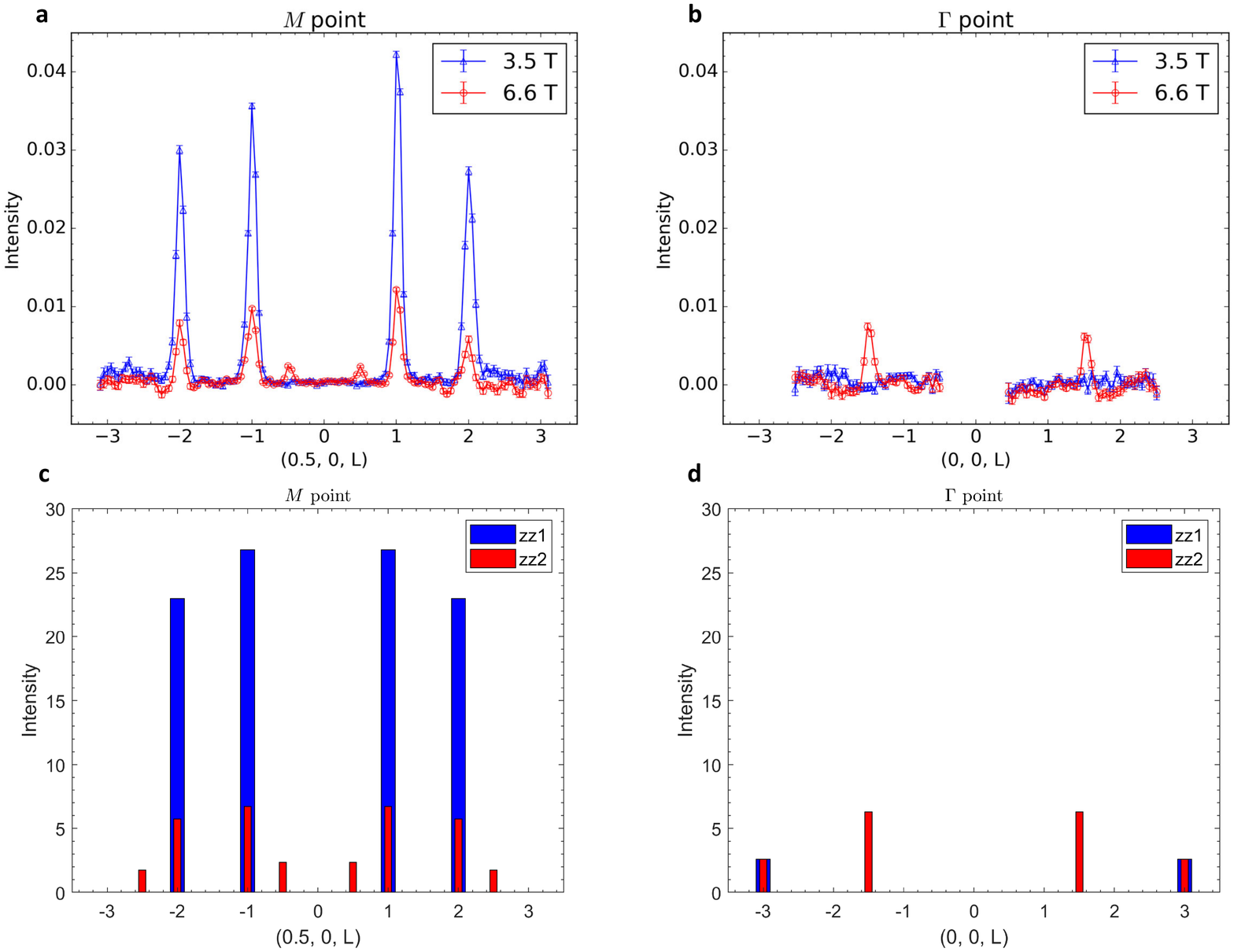}
\caption{{(a)} $\mathbf M$-point $(1/2,0,L)$ neutron diffraction intensities as a function of the out-of-plane wavevector component $L$ for $B=3.5$\,T (blue) and $B=6.6$\,T (red). The intensities have been averaged over the four different $\mathbf M$ points shown in Fig.~\ref{fig:diff_slices}.
{(b)} $\boldsymbol{\Gamma}$-point neutron diffraction intensities. The data in {(a,b)} are integrated in $\Delta H=\pm0.03$\,rlu and $\Delta K=\pm0.03$\,rlu.
{(c)} Magnetic structure factor modeling for the $\mathbf M$ point for the {\zza} (blue) and {\zzb} (red) phases as discussed in the text. Note that the intensities in the {\zzb} phase at integer $L$ are from the {\zzthree} and the ones at half-integer $L$ from the {\zzsix} structure which coexist at intermediate fields.
{(d)} Magnetic structure factor modeling for the $\boldsymbol{\Gamma}$ point (see text).
}
\label{fig:diff_cuts}
\end{figure*}

\section{Neutron diffraction}
\label{sec:neutron}

In order to characterize the intermediate-field phase, neutron diffraction data were taken in fields $\mathbf B\parallel\{1,1,0\}$, for which the two critical fields are separated the most, cf.\ Fig.~\ref{fig:phase_diagrams}{(b)}. In Fig.~\ref{fig:diff_slices} 2D slices of the honeycomb Brillouin zone for field strengths of 3.5\,T and 6.6\,T are shown. The perpendicular wavevector transfer along $L$ was integrated for narrow ranges around $L=1$ and $L=0.5$ respectively and averaged over positive and negative values. At 3.5\,T the intensity of the zig-zag magnetic Bragg peaks that remain above the domain re-population field of 2\,T was found to be the strongest \cite{Ban18}. The field of 6.6\,T is centered in the intermediate phase in between $B_\mathrm{c1}$ and $B_\mathrm{c2}$. Fig.~\ref{fig:diff_slices}{(a)} shows the four $L=1$ $\mathbf M$-point Bragg peaks at 3.5\,T. These four peaks also appear in the intermediate phase at 6.6\,T and remain at commensurate positions, as shown in Fig.~\ref{fig:diff_slices}{(b)} and in one-dimensional (1D) cuts within the honeycomb plane in the appendix. The most striking feature of the intermediate phase is the appearance of new zigzag Bragg peaks at half-integer values of $L$, as shown in Figs.~\ref{fig:diff_slices}{(d)} and \ref{fig:diff_cuts}{(a)}. No intensity was observed at these $L$ values at lower fields, see Fig.~\ref{fig:diff_slices}{(c)} for the same slice at 3.5\,T.

Since the intermediate phase is also characterized by Bragg peaks of the zigzag structure, we chose the naming convention {\zza} and {\zzb} for the two ordered phases. Most importantly, the phase transition represents a change of the 3D magnetic structure as indicated by Bragg peaks appearing at different values of $L$, but at the same positions within the honeycomb plane. The 3D character of the magnetic exchange interactions in {\rucl} has been discussed already in Refs.~\cite{Bal19,Jan20} and is confirmed by the observation of this transition.

To explore the nature of the {\zzb} phase in more detail, we show 1D cuts along the out-of-plane wavevector transfer $L$ in Fig.~\ref{fig:diff_cuts}{(a,b)}. The $\mathbf M$-point intensities in the {\zza} phase appear at values of $L=\pm1,\pm2$, consistent with the three-layer stacking of the crystal structure in the $R\bar{3}$ space group and an obverse-reverse twining ratio of approximately 50\%, as observed in large single-crystals of {\rucl} \cite{Par16,Cao21}. The Bragg peaks with $L=\pm2$ appear weaker because of (i) the magnetic form factor and (ii) the neutron polarization factor, which allows only the magnetic moment component perpendicular to the wavevector transfer $\mathbf{Q}$ to be measured. At 6.6 T, within the {\zzb} phase, the $\mathbf M$-point Bragg peaks with $L=\pm1,\pm2$ lose intensity, which can be understood as a consequence of the destabilization of the zigzag order in the vicinity of the transition to the disordered high-field phase.
Most importantly, Fig.~\ref{fig:diff_cuts}(a) again demonstrates the appearance of the new $\mathbf M$-point Bragg peaks at $L=\pm0.5$ in the {\zzb} phase.
The cut along $L$ for the 2D $\boldsymbol{\Gamma}$ point $(0,0,L)$ in Fig.~\ref{fig:diff_cuts}(b) reveals another set of magnetic Bragg peaks appearing at $L=\pm1.5$ in the {\zzb} phase.

\begin{figure}
\includegraphics[width=0.35\textwidth]{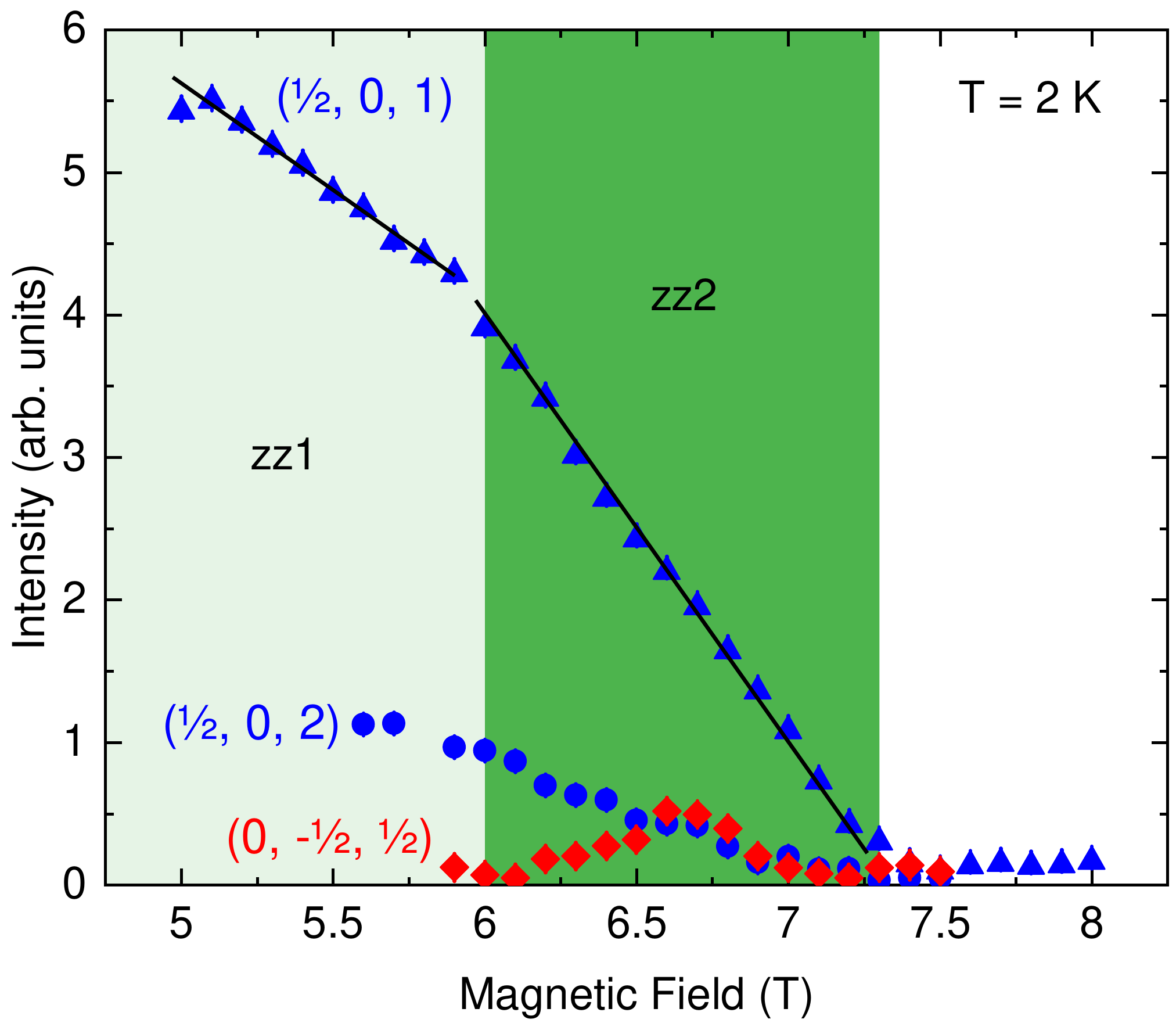}
\caption{Intensities of three different magnetic Bragg peaks at 2\,K as function of field strength for $\mathbf B \parallel \{1,1,0\}$, as obtained from Gaussian fits of 1D cuts through the data at each field strength. The lines are a guide to the eye. Error bars represent one standard deviation of the fitted intensity.
}
\label{fig:order_parameter}
\end{figure}

The magnetic-field dependence of the different sets of peak is presented in Fig.~\ref{fig:order_parameter}. At the transition to the {\zzb} phase, the intensity of peaks with integer $L$ exhibit a kink while the half-integer peaks start to emerge. The intensity of the latter goes through a maximum near 6.6\,T and at the transition to the disordered phase, all peaks simultaneously lose their intensities.

\section{Structure factor modeling}
\label{sec:structure_factor}

\begin{figure}
\includegraphics[width=\columnwidth]{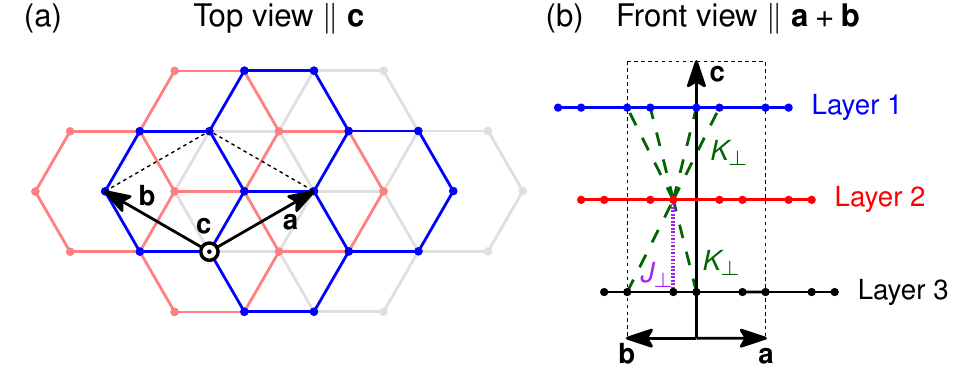}
\caption{$R\bar3$ crystal structure and interlayer couplings. The three inequivalent honeycomb layers are shown in black, red, and blue. The dashed rhombus indicates the crystallographic unit cell, consisting of two spins per layer. (a) Top view, with a viewpoint along the crystallographic $\mathbf c$ direction. (b) Front view, with a viewpoint along an in-plane direction perpendicular to a Ru-Ru bond. The interlayer couplings $J_\perp$ and $K_\perp$ are depicted in dashed green and dotted purple lines, respectively.
}
\label{fig:crystal}
\end{figure}

In order to understand the {\zzb} phase, we performed magnetic structure factor modeling in the $R\bar{3}$ space group for a zigzag magnetic structure on the honeycomb lattice. The $R\bar{3}$ crystal structure is illustrated in Fig.~\ref{fig:crystal}. The ordered moments are chosen to lie perpendicular to a Ru-Ru bond in the honeycomb plane and the angle between the ordered moments and the honeycomb plane was fixed to 15{\textdegree}, as refined from polarized and unpolarized single-crystal neutron diffraction data \cite{Cao21}. However, with the exception of the relative intensity of the $L=\pm2$ peaks, the results do not depend on this angle. Two different stackings of the in-plane zigzag configurations along the out-of-plane direction are possible, as illustrated in Fig.~\ref{fig:stackings}. The first one is characterized by an antiferromagnetic configuration between nearest interlayer neighbors and is compatible with the threefold stacking of the $R\bar{3}$ crystal structure ({\zzthree}). All Bragg peak intensities in the {\zza} phase are correctly reproduced by the {\zzthree} stacking, see the blue intensities in Fig.~\ref{fig:diff_cuts}{(c)}. The second one is characterized by a ferromagnetic alignment between nearest interlayer neighbors. This doubles the magnetic unit cell in the out-of-plane direction and leads to a sixfold zigzag stacking ({\zzsix}) as shown in Fig.~\ref{fig:stackings}(b). This structure consequently leads to magnetic intensities at half-integer positions along $L$, which in the $R\bar{3}$ space group appear at $L=\pm0.5,\pm2.5$. The observation of coexisting integer and half-integer peaks in Fig.~\ref{fig:diff_cuts}(a) leads us to assume phase coexistence of the {\zzthree} and {\zzsix} structures in the narrow field regime between 6 and 7.3\,T, representing the {\zzb} phase. This assumption is supported by our microscopic model below, which characterizes the transition at 6\,T as first-order, implying hysteresis effects.
The volume fractions of the {\zzthree} and {\zzsix} structures in the {\zzb} phase can be obtained from the field-dependent peak intensities within the first Brillouin zone. As visible in Fig.~\ref{fig:order_parameter} the intensity of the (0,-1/2,1/2) peak goes through a maximum at 6.6\,T and at this field value the {\zzsix} structure accounts for 1/4 of the ordered moment while 3/4 is still ordered in the {\zzthree} structure. It is important to note that the {\rucl} crystal was zero-field cooled for this experiment and the field was gradually increased at the base temperature of 2\,K.

In order to model the $\mathbf M$-point intensities in the {\zzb} phase at 6.6~T (Fig.~\ref{fig:diff_cuts}(a)), the magnetic structure factor is calculated for a superposition of the {\zzthree} and {\zzsix} structures with a ratio of 3/4 to 1/4. For comparison with the experimental data, the overall intensity is reduced by a factor of 4 compared to the {\zza} phase calculation at 3.5~T which is explained by the overall decrease of the order parameter. The resulting structure factor in Fig.~\ref{fig:diff_cuts}(c) agrees well with the observed intensities. From the experimental data, it is not clear whether intensity is present at $L=\pm2.5$, since the expected signal is within the level of the experimental noise. The $\boldsymbol{\Gamma}$-point intensities from Fig.~\ref{fig:diff_cuts}(b) are modeled in Fig.~\ref{fig:diff_cuts}(d). Two additional effects of the external field on the ordered structure are included. First, a 10\% uniform moment in the honeycomb planes along the field direction, which produces the magnetic peaks at $L=\pm3$. These were unobservable in the experiment, since they lie on top of intense nuclear peaks. This net ferromagnetic moment is independent of the 3D magnetic structure.
Second, a 10\% AFM moment perpendicular to the field direction that is uniform in each plane, but staggered between neighboring planes. This AFM moment has different effects on the magnetic structure factor in the {\zzthree} and {\zzsix} configurations. In the three-layer structure, this AFM moment cancels out completely and does not contribute to the structure factor. In the six-layer structure, it causes additional magnetic peaks at half the $L$ value of the nuclear peaks. In $R\bar{3}$, these magnetic peaks appear at $(0,0,\pm1.5)$ consistent with the observation from the experiment.
In sum, we were able to account for all observed magnetic Bragg peaks with a simple magnetic structure factor model based on two different stackings of the in-plane zigzag configurations.

\section{Microscopic spin model}
\label{sec:model}

In this section, we aim at constructing a microscopic spin model that describes the experimental findings. As the experiments indicate a field-driven change in the 3D magnetic structure, this requires a model involving interlayer couplings. Here, we construct such a model and show that it displays a first-order transition from {\zzthree} to {\zzsix} order for appropriately chosen parameters.

\subsection{3D spin model}

\begin{figure}
\includegraphics[width=\columnwidth]{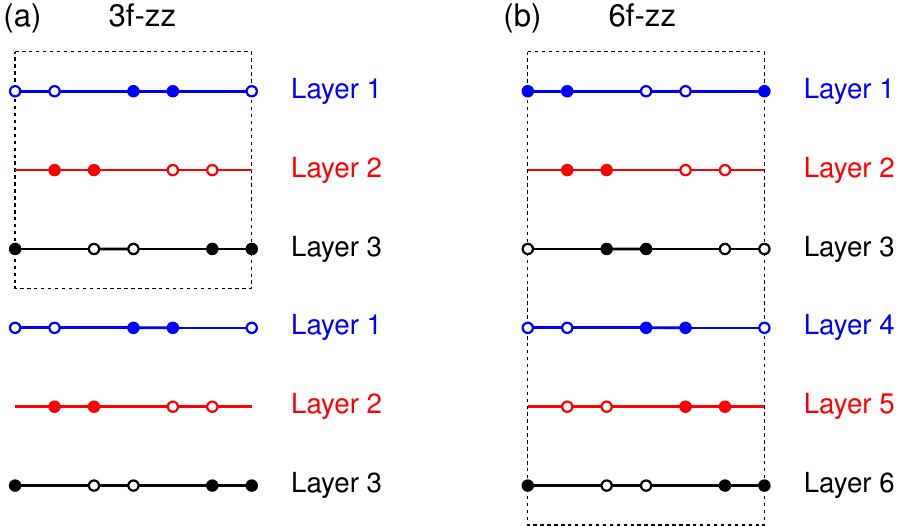}
\caption{
Two different stackings of in-plane zigzag configurations. The view is along the \textbf{a}+\textbf{b} direction as in Fig.~\ref{fig:crystal}(b) (perpendicular to a Ru-Ru bond). Spins that point into the same direction are represented by the same (filled or open) symbol; the different colors denote crystallographically inequivalent layers. The magnetic unit cells are indicated by dashed rectangles. (a) Threefold zigzag stacking (\zzthree). (b) Sixfold zigzag stacking (\zzsix). Note that in (a) spins directly above each other point in different directions (antiferromagnetic alignment) while they point in the same direction (ferromagnetic alignment) in (b).}
\label{fig:stackings}
\end{figure}

Restricting our attention to low temperatures, we assume a rhombohedral crystal structure with an $R\bar3$ space group, implying an undistorted $C_3^*$ symmetry \cite{Par16,Gla17}. The unit cell contains three honeycomb layers with two spins each, see Fig.~\ref{fig:crystal}.

Each spin has one interlayer neighbor that is located either right above or below it, depending on the sublattice index. The strictly vertical spin-spin exchange interaction is compatible with the $C_3^*$ symmetry only for a Heisenberg coupling, denoted as $J_\perp$ in Fig.~\ref{fig:crystal}(b). On the level of the next-nearest interlayer neighbors, bond-dependent interactions, such as a Kitaev coupling $K_\perp$ or an off-diagonal $\Gamma_\perp$, become symmetry allowed. We note that the nine next-nearest interlayer neighbors of each spin fall into two classes (with six and three members, respectively) that are distinguished by the presence or absence of a nearest-neighbor intralayer bond in one of the participating layers \cite{Jan20}. In what follows, we will not distinguish between these different next-nearest interlayer neighbors for simplicity.

As noted above, assuming a zigzag magnetic pattern within the honeycomb layers allows two different stackings in the out-of-plane direction, see Fig.~\ref{fig:stackings}.
$J_\perp > 0$ ($J_\perp < 0$) favors threefold (sixfold) stacking, respectively, independent of the particular spin directions. In order to assess the possibility of a field-induced transition between the different zigzag stackings, it is therefore mandatory to take spin-anisotropic interlayer couplings into account, as are symmetry-allowed for the next-nearest-neighbor interlayer bonds.
Here, we consider a simple model with the two interlayer couplings only. A Heisenberg nearest-neighbor interlayer coupling $J_\perp$ and a Kitaev next-nearest interlayer neighbor coupling $K_\perp$. The Hamiltonian may then be written as
\begin{align}
\label{eq:H}
	\mathcal{H} & =
	\sum_{n} \biggl\{
	\sum_{\langle ij\rangle_\gamma} \Bigl[
	J_1 \mathbf S_{n,i} \cdot \mathbf S_{n,j}  + K_1 S_{n,i}^\gamma S_{n,j}^\gamma 
	\notag\\ & \quad 
	+ \Gamma_1 \left( S_{n,i}^\alpha S_{n,j}^\beta + S_{n,i}^\beta S_{n,j}^\alpha\right)
	\Bigr]
	+ \sum_{\lllangle ij\rrrangle}
	J_3 \mathbf S_{n,i} \cdot \mathbf S_{n,j}
	\biggr\}
	\displaybreak[1] \nonumber \\ &\quad
	+ J_\perp \sum_{\langle ni,mi\rangle} \mathbf S_{n,i} \cdot \mathbf S_{m,i}
%
%
	+ K_\perp \sum_{\llangle ni,mj\rrangle_\gamma} S_{n,i}^\gamma S_{m,j}^\gamma
	\displaybreak[1] \nonumber \\ &\quad
	- \mu_\mathrm{B}\mathbf B \cdot g \sum_{ni} \mathbf S_{n,i},
\end{align}
where the indices $n,m$ label the layers and $i,j$ the sites within a given layer. In the above equation, the first two lines correspond to the usual intralayer interaction \cite{Win16,Win17,Jan17}, while the third line denotes the interlayer interactions. The fourth line is the Zeeman term for a uniform magnetic field $\mathbf B$, with $\mu_\mathrm{B}$ the Bohr magneton. We assume a diagonal $g$ tensor, $g = {\rm diag}(g_{ab}, g_{ab}, g_{c})$ in the crystallographic $(\mathbf a, \mathbf b, \mathbf c)$ basis,
with isotropic in-plane elements $g_a = g_b \equiv g_{ab}$, consistent with $C_3^*$ symmetry.
For the intralayer interactions, we use \cite{Win17,Jan17,Win18}
\begin{align} \label{eq:intralayer}
	(J_1, K_1, \Gamma_1, J_3) = (-0.1, -1, 0.5, 0.1) A,
\end{align}
where $A>0$ sets the overall energy scale.
Within a purely 2D modeling, this set of intralayer couplings fits well various experiments \cite{Ban17, Ban18, Wol17, Wang17, Gas20, Bac20}, but might require modifications upon the inclusion of sizable interlayer couplings \cite{Jan20}.
In order to constrain the parameter space, we hence assume for simplicity that both interlayer couplings are much smaller than the intralayer couplings, $|J_\perp|, |K_\perp| \ll A$.
Apart from the individual signs of $J_\perp$ and $K_\perp$, which will be constrained below, this leaves us with a single free parameter in our model, corresponding to the ratio $J_\perp/K_\perp$.
This turns out to be sufficient to describe well the qualitative features of the experiment.

\subsection{Phase diagram}

In the limit of $|J_\perp|, |K_\perp| \ll A$, the classical ground state of $\mathcal H$ can be found by minimizing the energy within each layer first and then considering the coupling between the layers as a perturbation. The ground state of the system with $J_\perp = K_\perp = 0$ is a zigzag pattern in each layer with the different stackings being degenerate. Small, but finite, $J_\perp$ and $K_\perp$ lift the degeneracy and stabilize either {\zzthree} or {\zzsix}, depending on the signs and relative sizes of the interlayer couplings. 
Previous modeling of the magnon spectrum at high fields~\cite{Jan20} suggested antiferromagnetic $J_\perp > 0$. Antiferromagnetic $J_\perp$ favors the {\zzthree} configuration. This configuration is in agreement with the measurements at low field in the {\zza} phase of {\rucl}, as discussed above. 
At intermediate fields between $B_\mathrm{c1}$ and $B_\mathrm{c2}$ in the {\zzb} phase of {\rucl}, the structure factor modeling indicated a coexistence of {\zzthree} and {\zzsix}. The change of the zigzag stackings as a function of field suggests an exchange frustration mechanism arising from the interlayer couplings.
With antiferromagnetic $J_\perp > 0$, such an interlayer frustration can be achieved by assuming ferromagnetic $K_\perp < 0$.
In fact, using this sign structure of the two interlayer couplings and appropriate chosen magnitudes, we indeed find a transition between a {\zzthree} configuration at low fields and a {\zzsix} configuration at intermediate fields, before the zigzag order is completely destabilized at the transition to the disordered phase, in agreement with the experiment.
Within our simple interlayer model and for the present set of intralayer couplings [Eq.~\eqref{eq:intralayer}], this occurs for $\mathbf B \parallel \{1,1,0\}$ within a narrow parameter range $1.1165 < (-K_\perp)/J_\perp < 1.1489$.
The emergence of this field-induced transition between the different zigzag stackings can be understood as a consequence of the inhomogeneous canting of the spins for $\mathbf B \parallel \{1,1,0\}$, which leads to a different dependence on the canting angles of the energies of the {\zzthree} and {\zzsix} states.

By contrast, for $\mathbf B \parallel \{1,0,0\}$, the canting is homogenous and the difference between the {\zzthree} and {\zzsix} energies can be written as
\begin{align}
	\frac{\Delta E}{N S^2} &=
	\left[- J_\perp + K_\perp \left(\cos^2 \theta - \sqrt{2} \sin 2 \theta \right) \right] 
	\frac{1-\cos 2\vartheta_B}{2},
\end{align}
where $\theta \equiv \theta(\Gamma_1/K_1) \in [-\arctan(1/\sqrt{2}),0]$ parametrizes the direction of the spins at zero field \cite{Jan17} and $\vartheta_B \equiv \vartheta(B) = \angle(\mathbf S_i, \mathbf B) \in (0,\pi/2]$ is the homogeneous canting angle. $N$ is the total number of spins and $S = |\mathbf S| = 1/2$ for \rucl. For fixed couplings, the energy difference is therefore always either positive or negative, but cannot change sign as a function of the field strength $B$. For $\mathbf B \parallel \{1,0,0\}$, a field-induced transition between {\zzthree} and {\zzsix} is therefore not possible within our classical model and for infinitesimal interlayer couplings.  We have checked that this remains true when small off-diagonal interlayer couplings $\Gamma_\perp$ are taken into account.

\begin{figure}
\includegraphics[width=\columnwidth]{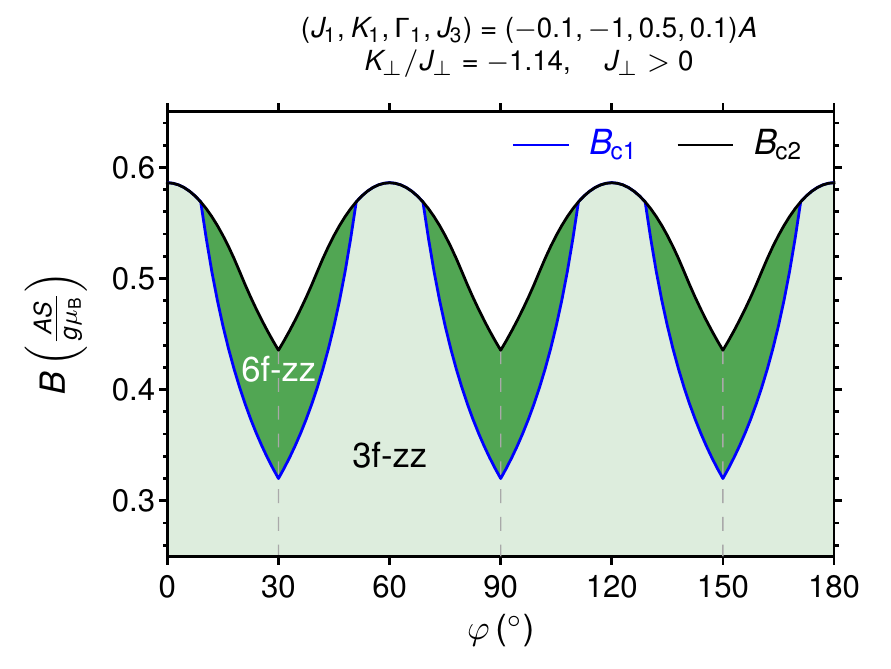}
\caption{
Classical phase diagram of microscopic spin model as function of magnetic field $B$ and in-plane angle $\varphi$ in the limit of small interlayer couplings, $|K_\perp|, |J_\perp| \ll A$, with $(-K_\perp) / J_\perp = 1.14$. {\zzthree} and {\zzsix} denote ordered zigzag configurations with threefold and sixfold layer periodicity, respectively. The experimental phases {\zza} and {\zzb} in Fig.~\ref{fig:phase_diagrams} are described by {\zzthree} and a coexistence of {\zzthree} and {\zzsix}, respectively. $B_\mathrm{c1}$ (blue) denotes the first-order transition between different zigzag stackings, $B_\mathrm{c2}$ (black) denotes the transition to the disordered phase.}
\label{fig:phasediag-th}
\end{figure}

The full classical phase diagram of our microscopic model as a function of the in-plane angle $\varphi$ is shown in Fig.~\ref{fig:phasediag-th}. Here, we have chosen $(-K_\perp) / J_\perp = 1.14$ with $J_\perp > 0$. We recall that we work in the limit $|J_\perp|, |K_\perp| \ll A$, where the phase diagram only depends on the ratio of $J_\perp$ and $K_\perp$ and not on their individual magnitudes.
The phase diagram illustrates that the threefold zigzag stacking ({\zzthree}) is stabilized for small fields, while a transition to an intermediate ordered phase with sixfold zigzag stacking ({\zzsix}) is found at elevated in-plane fields perpendicular to Ru-Ru bonds.
The minimal microscopic theory presented here gives a satisfactory explanation for the observed features in the {\zza} and {\zzb} phases with one exception: It does not predict the small inter-layer staggered antiferromagnetic component necessary to produce the peaks seen in the {\zzb} phase at $(0,0,\pm1.5)$.
Importantly, the transition at $B_\mathrm{c1}$ is first-order, such that coexistence of the {\zzthree} and {\zzsix} configurations, and therefore their Bragg peaks, can be naturally expected in its vicinity as a consequence of hysteresis effects.
We note that a first-order transition from {\zza} to {\zzb} at $B_\mathrm{c1}$ is consistent with other thermodynamic measurements reported in the literature \cite{Bac20,Sch20,Bac21}.

\subsection{Magnetization curves}

\begin{figure}
	\includegraphics[width=\columnwidth]{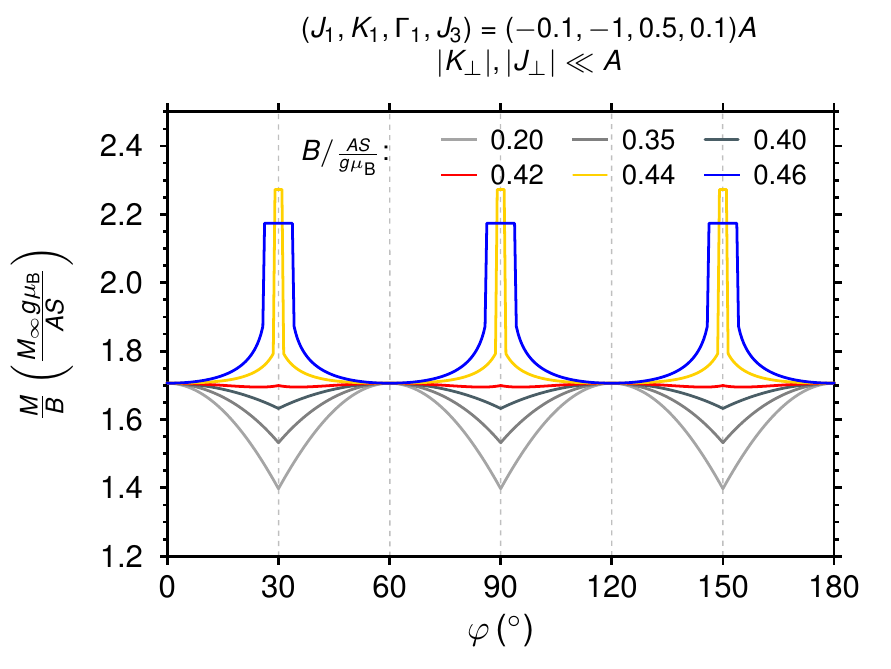}
	\caption{\label{fig:magnetization-classical}%
	Classical magnetization $(M/B)$ at various fields in the microscopic spin model as function of in-plane angle $\varphi$ in the limit of small interlayer couplings, $|K_\perp|, |J_\perp| \ll A$, where $A$ sets the overall energy scale.}
\end{figure}

The magnetization at fixed field strengths as a function of the in-plane angle $\varphi$ is shown in Fig.~\ref{fig:magnetization-classical}. The low-field limit of this magnetization curve has previously been discussed in Ref.~\cite{Jan17}. In this limit, the magnetization is maximal for fields along Ru-Ru bonds and minimal for fields perpendicular to Ru-Ru bonds, with a characteristic kink that can be understood as a domain switching effect. 
Increasing the field strength now shifts these minima upwards, such that the magnetization becomes nearly $\varphi$ independent for fields just below $B_\mathrm{c2}(30^\circ)$, until eventually the maxima and minima exchange for fields between $B_\mathrm{c2}(30^\circ)$ and $B_\mathrm{c2}(0^\circ)$, in qualitative agreement with the experiment.
In the polarized phase, the magnetization curve is flat, which is a classical property that will change upon the inclusion of quantum fluctuations~\cite{Jan17}.

\section{Discussion}
\label{sec:disc}

The comparison of the experimental and theoretical results in terms of the angle-dependent phase diagrams [Figs.~\ref{fig:phase_diagrams}(b) and \ref{fig:phasediag-th}], as well as the magnetization curves [Figs.~\ref{fig:bulk_properties}(g) and \ref{fig:magnetization-classical}], reveals that the key features of the experiment are reproduced by the model calculation:
(i)~A first-order phase transition occurs at a critical field $B_\mathrm{c1}$ between zigzag structures with threefold and sixfold stacking.
(ii)~Both the critical fields and the magnetization curves have (approximately) a 60$^\circ$ periodicity, consistent with the $C_3$ rotational symmetry of the honeycomb lattice.
(iii)~The critical fields are maximal for fields along Ru-Ru bonds ($\varphi \equiv 0^\circ \bmod 60^\circ$) and minimal for fields perpendicular to a bond ($\varphi \equiv 30^\circ \bmod 60^\circ$).
(iv)~The width of the intermediate phase is maximal when the critical field is minimal, and vice versa.
(v)~For fixed low fields, the magnetization $M$ is maximal for $\varphi \equiv 0^\circ \bmod 60^\circ$ and has kink-like minima at $\varphi \equiv 30^\circ \bmod 60^\circ$. 
(vi)~At fixed elevated fields close to the transition to the disordered phase, on the other hand, maxima and minima in the magnetization as function of $\varphi$ exchange.

\section{Conclusions}
\label{sec:conclusions}

We have demonstrated the existence of a field-induced intermediate ordered phase in {\rucl} at fields just below the field where the magnetic order is found to be suppressed completely.
This phase is characterized by an in-plane zigzag configuration with a stacking periodicity in the out-of-plane direction that is doubled in comparison with the low-field zigzag phase. 
The fact that the transition at $B_\mathrm{c1}$ involves a change of the 3D magnetic structure shows that interlayer interactions are important in {\rucl} and should be included in the minimal model Hamiltonian \cite{Jan20}. 
We have proposed a simple model including Heisenberg as well as Kitaev interlayer interactions that describes the two different zigzag stackings, their field-induced transition, as well as the magnetization measurements qualitatively well.
We emphasize, however, that this analysis does not enable us to draw conclusions concerning the absolute value of the interlayer interactions in {\rucl}.
The previous 3D modeling \cite{Jan20} of the out-of-plane neutron scattering data \cite{Bal19} suggested a nearest-neighbor interlayer coupling of the order of $J_\perp \lesssim 1$\,meV. The present results indicate that bond-dependent interlayer interactions of the same order may play an equally important role. While such interactions are allowed by symmetry and therefore likely to be present, a microscopic mechanism that explains their significance is currently not known. 

Finally, we note that the data presented here are not of sufficient resolution at high fields to comment on potential additional phase transitions in the disordered regime \cite{Kas18,Bal19,Wul19,Yokoi20}. The ac susceptibility data appears mostly featureless above 8\,T, however, an additional small kink can be seen between 8 and 10\,T in Fig.~\ref{fig:bulk_properties}{(e)}, also visible in the false color plot in Fig.~\ref{fig:phase_diagrams}{(a)}. This might be related to the topological transition out of a potential QSL phase \cite{Kas18,Bal19}.


\acknowledgments

We thank E. C. Andrade, B. B\"uchner, P. M. C\^{o}nsoli, S. Koch, S. Rachel, and A. U. B. Wolter
for illuminating discussions and collaborations on related work.
This research used resources at the Spallation Neutron Source, a DOE Office of Science User Facility operated by the Oak Ridge National Laboratory.
CB, AB, YHL, and SEN were supported by the Division of Scientific User Facilities, Basic Energy Sciences US DOE, PLK and DGM by the Gordon and Betty Moore Foundation's EPiQS Initiative through Grant GBMF4416, JQY by the U.S. Department of Energy, Office of Science, Office of Basic Energy Sciences, Materials Sciences and Engineering Division. AB and SEN acknowledges support by the Quantum Science Center (QSC), a National Quantum Information Science Research Center of the U.S. Department of Energy (DOE). 
LJ and MV acknowledge financial support from the Deutsche Forschungsgemeinschaft (DFG) through SFB 1143 (project id 247310070) and the W\"urzburg-Dresden Cluster of Excellence on Complexity and Topology in Quantum Matter -- \textit{ct.qmat} (EXC 2147, project id 390858490). The work of LJ is funded by the DFG through the Emmy Noether program (JA2306/4-1, project id 411750675).


\appendix*

\section*{Appendix: Detailed cuts through the neutron diffraction data}\label{app:cuts}

\begin{figure*}
\includegraphics[width=\textwidth]{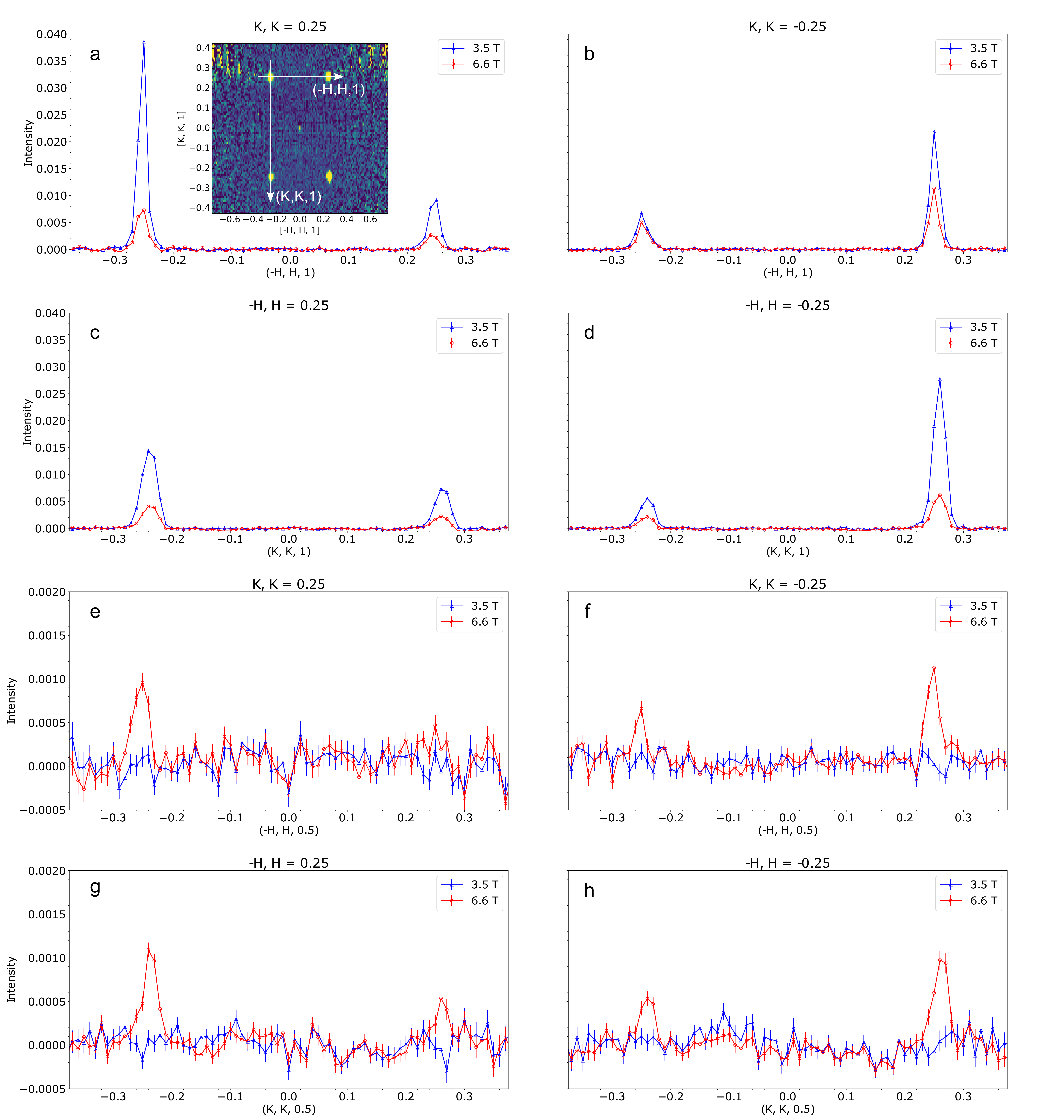}
\caption{1D cuts through the neutron diffraction data in the honeycomb plane at 3.5 and 6.6\,T. Panels a-d  show cuts with $|L|=1$ and e-f with $|L|=0.5$. The data is integrated over the perpendicular directions by $\pm0.03$\,r.l.u. in the honeycomb plane and by $\pm0.025$\,r.l.u. along $L$. In addition the data is averaged over positive and negative $L$. The inset of panel a shows the direction of the cuts in the honeycomb plane. Note the difference in $y$-axes scaling between panels a-d and e-h. 
}
\label{fig:detailed_cuts}
\end{figure*}

Fig. \ref{fig:detailed_cuts} shows cuts through the $\mathbf M$ points taken within the honeycomb plane. Each of the four $\mathbf M$ points $(0.5,0,L)$, $(0,0.5,L)$, $(0,-0.5,L)$, and $(-0.5,0,L)$ is cut in two different directions indicated in the inset of Fig.~\ref{fig:detailed_cuts}(a). While panels (a), (b), (e), and (f) are cuts along $(-H,H)$, panels (c), (d), (g), and (h) are cuts along $(K,K)$.

Panels (a-d), which are taken at integer $L$, show that the magnetic Bragg peaks remain at commensurate positions upon entering the {\zzb} phase (red data points). The only difference between 3.5\,T and 6.6\,T is a reduction in intensity. This agrees with the expectation for an overall order parameter upon approaching the point at which magnetic order is destroyed, cf.\ Fig.~\ref{fig:order_parameter}. The difference in intensities between the four $\mathbf M$ points, which also changes under increasing magnetic field, is attributed to the presence of different crystallographic and magnetic domains and their nontrivial field evolutions.

Panels (e-h) show the appearance of new magnetic peaks in the {\zzb} phase (red data points) at half-integer $L$ values, which are approximately one order of magnitude weaker compared to the ones at integer $L$. These also appear at commensurate positions and again show differences in intensities between the four $\mathbf M$ points. Since they are absent in the {\zza} phase (blue data points), their appearance clearly marks the entrance into a new thermodynamic phase.


\begin{thebibliography}{99}
\bibitem{prev_note}
The present paper supersedes Ref.~\cite{Kel18b}. That reference proposed that the intermediate-field ordered phase displays an in-plane magnetic order different from zigzag, which was disproven by subsequent neutron diffraction measurements reported here. The model calculations in Ref.~\cite{Kel18b} are correct, but do not apply to \rucl.
\end{thebibliography}
\end{document}